\documentclass[9pt, conference]{IEEEtran}

\IEEEoverridecommandlockouts
\usepackage{cite}
\usepackage{amsmath,amssymb,amsfonts}
\usepackage{algorithmic}
\usepackage{comment}
\usepackage{graphicx}
\usepackage{textcomp}
\usepackage{xcolor}
\usepackage{booktabs}
\def\BibTeX{{\rm B\kern-.05em{\sc i\kern-.025em b}\kern-.08em
    T\kern-.1667em\lower.7ex\hbox{E}\kern-.125emX}}

\begin{document}

\title{A decade of DCASE: Achievements, practices, evaluations and future challenges
\thanks{This work was supported in part by Academy of Finland grant 332063 ``Teaching machines to listen",
and by 
the UK Engineering and Physical Sciences Research Council (EPSRC) under Grant EP/T019751/1 ``AI for Sound''.
For the purpose of open access, the authors have applied a
Creative Commons Attribution (CC BY) license to any Author
Accepted Manuscript version arising.
This publication is supported by datasets that
are openly available at locations referenced in this paper.
}
}

\author{
\IEEEauthorblockN{Annamaria Mesaros$^1$, Romain Serizel$^2$, Toni Heittola$^1$, Tuomas Virtanen$^1$, Mark D. Plumbley$^3$}
\IEEEauthorblockA{
\textit{$^1$ Signal Processing Research Center, Tampere University}, Tampere, Finland \\
\textit{$^2$ Université de Lorraine, CNRS, Inria, Loria}, Nancy, France \\
\textit{$^3$ Centre for Vision, Speech and Signal Processing, University of Surrey, Guildford, UK}
}
annamaria.mesaros@tuni.fi, romain.serizel@loria.fr , toni.heittola@tuni.fi, m.plumbley@surrey.ac.uk , tuomas.virtanen@tuni.fi
}

\maketitle

\begin{abstract}
This paper introduces briefly the history and growth of the Detection and Classification of Acoustic Scenes and Events (DCASE) challenge, workshop, research area and research community. Created in 2013 as a data evaluation challenge, DCASE has become a major research topic in the Audio and Acoustic Signal Processing area. Its success comes from a combination of factors: the challenge offers a large variety of tasks that are renewed each year; and the workshop offers a channel for dissemination of related work, engaging a young and dynamic community. At the same time, DCASE faces its own challenges, growing and expanding to different areas. 
One of the core principles of DCASE is open science and reproducibility: publicly available datasets, baseline systems, technical reports and workshop publications. 
While the DCASE challenge and workshop are independent of IEEE SPS, the challenge receives annual endorsement from the AASP TC, and the DCASE community contributes significantly to the ICASSP flagship conference and the success of SPS in many of its activities.
\end{abstract}


\section{Introduction}
\label{sec:intro}
Early research in the field of acoustic scene analysis and event detection comprises work on computational auditory scene analysis (CASA) which aimed to mimic human auditory perception to segregate and identify sound sources in complex audio environments \cite{ellis1996prediction}, \cite{rosenthal2021computational}. Publications focused on various aspects of audio signal processing, such as feature extraction, and pattern recognition algorithms for sound classification, often comparing effectiveness of known features from other domains, like mel-frequency cepstral coefficients (MFCCs) and temporal features like zero crossing rate, spectral centroid, short-time energy for classifying environmental sounds \cite{Chu2009} or scenes \cite{eronen2006audio}. Classification techniques were based mainly on support vector machines (SVMs) \cite{Chu2009, eronen2006audio}, Gaussian mixture models and hidden Markov models \cite{eronen2006audio}. 

Prior to the DCASE challenges, there were limited publicly available datasets for training and evaluating acoustic scene or sound recognition systems, most notably RWCP \cite{nakamura2000acoustical}. In terms of challenges, the CLEAR Evaluations on the Classification of Events, Activities, and Relationships conducted in 2006 and 2007 had a task focused on Acoustic Scene Analysis to detect 12 categories of \textit{non-speech noises}; the task had 2 entries in 2006 \cite{CLEAR2006}, and 5 in 2007 \cite{CLEAR2007}, but there was no further activity after the second edition. 
Subsequently, studies on environmental sound recognition used in-house datasets and were therefore non-reproducible. Moreover, the choice of evaluation procedures was also up to the author, with no clear agreement on metrics or standardized benchmarks. This lack of standardization hindered the comparability of results across different studies. 

The formalization of the DCASE challenges in 2013  \cite{Stowell2015} marked a significant change. With standardized datasets and evaluation protocols, the DCASE challenges provided a structured platform for researchers to benchmark their algorithms against a common dataset, leading to more reproducible and comparable results. The result was a more collaborative and competitive research environment that fostered significant advancements in the field for acoustic scene classification, sound event detection, and other related tasks.

\section{Open science through data challenges}
Looking back for a brief history of data evaluation challenges, MIREX \cite{Downie2010} or CHiME \cite{barker2018fifthchimespeechseparation} stand out. MIREX began in 2005 as an annual evaluation campaign for Music Information Retrieval (MIR), evaluating algorithms for tasks like music genre classification, melody extraction,  and music similarity. Over time, MIREX expanded to more complex tasks, becoming a benchmark for MIR research. MIREX promoted open-source practices and detailed reporting, emphasizing rigorous evaluation and reproducibility, even though it was not explicitly focused on open science as we know it today. 
The CHiME challenges began in 2011, focusing on robust automatic speech recognition (ASR) in noisy environments, with typical tasks being speech enhancement, speaker diarization and speech recognition in real-world noisy environments.  Through multiple iterations, CHiME has introduced more complex scenarios like distant microphone conversational speech recognition in everyday home environments \cite{barker2018fifthchimespeechseparation}.
DCASE started in 2013 under endorsement of the IEEE Audio and Acoustic Signal Processing (AASP) Technical Committee through its AASP Challenges Subcommittee\footnote{https://signalprocessingsociety.org/community-involvement/audio-and-acoustic-signal-processing/aasp-challenges}, and became a very successful and far-reaching endeavor. 

\subsection{The first DCASE Challenge}

Before the first DCASE challenge, a one-day “Machine Listening Workshop” organized at Queen Mary University of London, UK in December 2010 provided an early indication of the interest in research in this area. 
Subsequent discussions in 2011 between Queen Mary University of London, UK (Benetos, Giannoulis, Stowell, Plumbley) and IRCAM, Paris, France (Rossignol, Lagrange) led to the idea for the first challenge. At around that time, Plumbley had also joined the Audio and Acoustics Signal Processing Technical Committee (AASP TC) of the IEEE Signal Processing Society, which was organizing a new ``IEEE AASP Challenge'' series. A proposal for a challenge in ``Detection and Classification of Acoustic Scenes and Events'' was submitted and approved in early 2012, becoming the second ``IEEE AASP Challenge''. The Chairs of the following year’s WASPAA 2013 workshop also agreed to support the challenge, through a special poster session with an overview oral presentation.

The first DCASE challenge included two tasks: Acoustic scene classification (ASC), and Sound Event Detection (SED). These tasks have been present in each edition to date, but their setups have evolved to be more complex and closer to real-life applicability, and datasets became significantly larger and more diverse.
For the Acoustic Scene Classification task, after considering whether to use existing data, it was decided to collect new data, to avoid inconsistency of microphones and recording equipment. Recordings were made from ten different types of sound scenes in the London area, 
and were released in 30-second segments, with 10 examples of each of the 10 scenes, totalling 50 minutes of audio.
The Sound Event Detection task included two sub-tasks, based on sounds recorded in offices: an ``Office Live’’ (OL) task and an ``Office Synthetic’’ (OS) sub-task. The OL recordings were live recordings of scripted event sequences, with no overlapping events.
The OS recordings were synthesized from live recordings of individual events, combined into synthetic mixtures which included ambient background, which allowed different polyphony (overlap) of sound events and different signal-to-noise (SNR) levels of event sounds over background.

\subsection{DCASE Challenges 2016-2024}

The DCASE Challenge was restarted in 2016 with support from an ERC Starting Grant by Virtanen, which facilitated data collection and annotation at a larger scale. Initially organized by a core group of people through existing collaborations, it soon became evident that this approach is unsustainable due to its popularity, reflected in large number of participants. Thus, an open call for tasks was introduced.

The first open call for task proposals, published in December 2018, seeked to include new tasks, unrelated to those from the previous years. For 2019, the Steering Group selected five tasks; these included the longstanding ASC and SED, but also new tasks like audio tagging (AT) with noisy labels \cite{Fonseca2019_ICASSP}, sound event localization and detection (SELD) \cite{Politis2020}. 
The decentralized organization makes the challenge a community effort, distributing the organization effort to multiple researchers who have full control over their task. The challenge operates under unified coordination, with the challenge coordinators (Mesaros and Serizel) aligning the format of the tasks. 

The open call for task organization had a profound impact on the challenge and the entire research field, diversifying the topics of the challenge. New research directions that emerged are sound event localization and detection (SELD), the use of weakly-labeled data \cite{Turpault2019_DCASE}, and few-shot learning for bioacoustic event detection \cite{Nolasco2023}. Furthermore, the introduction of tasks like automated audio captioning \cite{Mei2022_captioning} and language-based audio retrieval \cite{Xie_2022_dcase} highlights the growing intersection of audio processing with natural language processing. The latest addition to the DCASE challenge is an audio generation task \cite{Choi2023}, marking a significant step forward from what was essentially audio signal processing and analysis. 
As a result, the DCASE research field has seen notable advancements, the diversity in tasks, shown in numbers in Table \ref{tab:DCASE_challenge_stats}, ensuring that it remains vibrant, highly-engaging, and continually evolving.

\begin{table}[]
    \caption{DCASE Challenge statistics.}
    \label{tab:DCASE_challenge_stats}
    \centering
    \begin{tabular}{c|cc|cc}
    \toprule
     & Research & Tasks and &  &  \\
    Edition & topics & Subtasks & Teams & Entries \\
    \midrule
    2023 & 7 & 9 & 123 & 428  \\
    2022 & 6 & 7 & 135 & 410 \\
    2021 & 6 & 8 & 127 & 393 \\
    2020 & 6 & 7 & 138 & 473 \\
    2019 & 5 & 7 & 109 & 311 \\
    2018 & 5 & 7 & 81 & 223 \\
    2017 & 4 & 4 & 74 & 200 \\
    2016 & 4 & 4 & 66 & 85 \\
    2013 & 2 & 2 & 21 & 31 \\
    \bottomrule
    \end{tabular}
    \vspace{-12pt}
\end{table}

\subsection{The DCASE Challenge process}

The typical timeline of the challenge starts with the open call for task proposals. The review and selection of tasks is done by the Steering Group, with the selected tasks being announced in January. Task organizers are responsible with producing and publishing the open datasets and baseline systems for their task, and evaluating the submissions. The challenge opens in March/April, when complete information on tasks is provided to potential participants: a detailed task description including the evaluation procedure, the development dataset, and the baseline system (code and results according to the task requirements). 

Participants have about two months to develop their algorithms. The evaluation datasets (when required by the task) are provided 2-4 weeks before the submission deadline. The participants are expected to submit the system outputs for the evaluation data, a technical report describing their method, and additional meta information related to the submission, using provided templates. The notable difference from the first DCASE is the submission of system outputs instead of code: running the code submitted by participants was a very time-consuming process \cite{Barchiesi2015}, hence the choice of simplifying the evaluation process. 

Because the key dates are aligned for all tasks, DCASE challenge feels like one single event, even though there are multiple teams working under its umbrella. Publication of task descriptions, development datasets, submission deadline and publication of results happens for all tasks at once (with small exceptions). The technical reports are published unaltered on the challenge website at the same time with the results.

\section{DCASE Workshops}

Dissemination of the challenge results was initially organized during a special session at WASPAA 2013, later as a satellite workshop to EUSIPCO 2016, where many participants attended only the DCASE workshop, not necessarily the main conference. This motivated the regular organisation of a standalone event. Since then, the workshop has been held every year as a companion event to the challenge.  Although the workshop and challenge teams operate independently, they maintain communication through the challenge coordinators to ensure their deadlines are compatible.

The DCASE workshop is now a two-and-a-half-day event that attracts around 100 participants each year. Approximately 50 scientific papers are published at each edition, with an acceptance rate of about 50\%. The workshop accounts for roughly 15\% of the annual publications on DCASE-related topics (included in Fig.~\ref{fig:paper_stats}), making it one of the leading venues for research in this area, alongside ICASSP. In line with the challenge’s principle of open science, the workshop proceedings are published under a Creative Commons license.

\begin{figure}[t]
\centerline{\includegraphics[width=1.0\columnwidth]{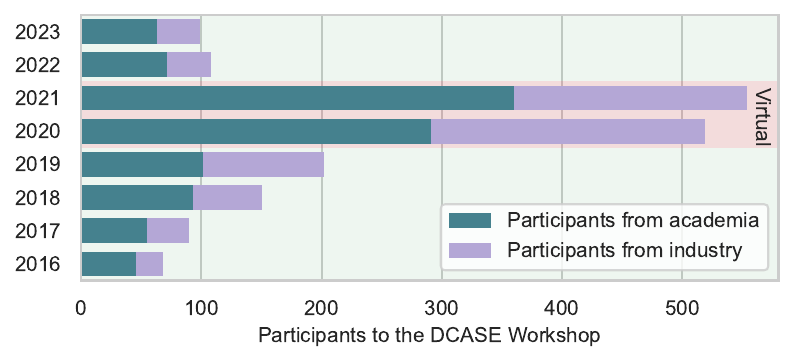}}
\vspace{-8pt}
    \caption{DCASE Challenge and Workshop participation rates (DCASE 2020 Workshop and DCASE 2021 Workshop were virtual and had free registration}
\label{fig:DCASE_Workshop_stats}
\vspace{-6pt}
\end{figure}

Due to the close relationship between the workshop and the challenge, the early editions of the workshop primarily featured papers related to challenge submissions, comprising about 80\% of the content until 2018. This trend has changed since 2019 and challenge-related papers now make up about half of the workshop’s publications. 
The number of challenge participants increased rapidly from 2013 to 2019 and has since stabilized, as reflected in Table \ref{tab:DCASE_challenge_stats}.
A similar trend is observed in workshop attendance  (Fig.~\ref{fig:DCASE_Workshop_stats}), with the exception of 2020 and 2021, when the workshop was held virtually due to the pandemic and waived registration fees for all participants. 
Interestingly, the DCASE workshop has attracted significant interest from industry since its inception, with 32\% of participants from industry in the first edition. This participation rate has remained stable, averaging 35-38\%, and even reaching 50\% in 2019.
This balance fosters insightful discussions on the relationship between academic research and real-life applications.

\section{DCASE as a research area}

\begin{figure}[t]
\centerline{\includegraphics[width=1.0\columnwidth]{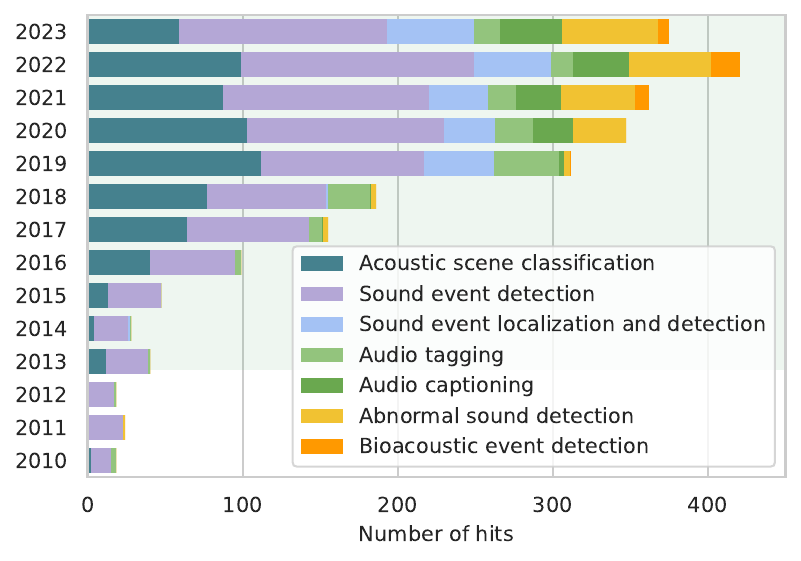}}
\vspace{-8pt}
    \caption{Number of publications related to DCASE Challenge tasks: Google Scholar hits for papers containing the respective keywords in their title}
\label{fig:paper_stats}
\vspace{-10pt}
\end{figure}

The DCASE research area began to flourish once public datasets and clearly defined research questions were introduced through the challenge tasks. The availability of datasets and baseline system code for each task provided researchers with a solid starting point in their work. Figure \ref{fig:paper_stats} shows the increase in scientific publications over the years that contain key terms from the DCASE challenge tasks in their title. The tasks within DCASE guide several key research directions, of which a few are briefly introduced below.

\textbf{Acoustic scene classification} aims to classify a test recording into one of the provided predefined classes that characterizes the environment in which it was recorded. Such acoustic scenes include "airport", "public square", and "metro". The ASC task has been active for all ten editions of the DCASE challenge, with changes in the classification setup and the dataset. From a dataset of 50 minutes and 10 classes in 2013 \cite{Barchiesi2015}, 13 hours and 15 classes in 2017 \cite{mesaros2017dcase}, the largest dataset to date is TAU Urban Acoustic Scenes 2019 \cite{Mesaros2018}, consisting of 40 hours of audio and 10 classes, simultaneously recorded with four devices. Research questions tackled multi-device data \cite{Heittola2020}, low-complexity \cite{Martin-Morato2022}, and data-efficient ASC \cite{Schmid2024}.

\textbf{Sound event detection} aims to identify sound event classes and their time boundaries in an audio recording. This task was also present in all the ten editions, with iterative changes. While the task setup was straightforward in the first editions~\cite{Stowell2015, mesaros2017dcase}, the 2017 edition of the challenge introduced the use of weakly-labeled training data to overcome the strongly labeled data shortage~\cite{Mesaros2019_TASLP}. Since then, the task has evolved to include synthetic training data~\cite{Turpault2019_DCASE}, sound separation as a pre-processing~\cite{Turpault2020b}, systematic reporting of the computational footprint~\cite{ronchini2024performance} and the use of soft labels~\cite{Martinmorato2023} in a framework with potentially missing labels during training~\cite{cornell2024dcase}.

\textbf{Sound event localization and detection} was introduced in 2019, aiming to jointly recognize and localize sound events in 3D space \cite{Adavanne2018_JSTSP}; the task involves detecting the temporal activities of sound events and estimating their spatial locations, using both first-order ambisonics and microphone array recordings. The latest development of the SELD task include using audio-visual input \cite{Shimada2023starss23} and, in 2024, distance estimation \cite{Krause2024}.

\textbf{Bioacoustic sound event detection}, a task introduced in 2022, focuses on few-shot learning, which requires algorithms to detect and classify sounds with only a few examples. This approach has evolved over the years, addressing the challenges of data sparsity and class imbalance in bioacoustic datasets \cite{Nolasco2023}. The task’s objective is to develop reliable algorithms capable of identifying animal vocalizations in noisy environments with minimal labeled data \cite{liang2024bioacoustic}.

\textbf{Anomalous sound detection} focuses on identifying whether a sound emitted from a machine is normal or anomalous, given that only normal sounds are provided for training \cite{Koizumi_DCASE2020_01}. The task was first introduced in 2020 and has gained popularity quickly due to its clear industrial applicability. The task has evolved to address practical challenges such as domain shifts \cite{Kawaguchi2021} and the need for rapid deployment in new environments \cite{Nishida_arXiv2024_01}. 

\textbf{Automated audio captioning} and \textbf{Language-based audio retrieval} are relatively new tasks. Automated audio captioning~\cite{Drossos_2017_waspaa} was introduced in 2021 as the first textual language based task in the DCASE challenge, supplemented the following year with the language-based audio retrieval task~\cite{Xie_2022_dcase}. The goal of these tasks is to link the audio and text modalities. Automated audio captioning aims to describe audio recordings in a natural language, beyond sound scene or event classes. Conversely, language-based audio retrieval focuses on finding audio recordings that correspond to a given  natural language description from a pool of recordings.

As exemplified above, DCASE has become a diverse research area, expanding from the initial audio analysis to encompass different modalities and tackle real-world applicability scenarios. In addition to the dedicated DCASE challenge and workshop, DCASE is strongly represented in the IEEE SPS conferences and journals; for example there were 65 DCASE-related papers in ICASSP 2024, a number similar to the size of a DCASE workshop. 

\section{DCASE Community practices and own challenges}

DCASE has been successful in building a thriving research community into which a significant number of volunteers are actively contributing. Several factors have enabled building the community:

1) Many researchers in the community shared backgrounds in related audio processing fields like music or speech processing, information retrieval, sound source separation, machine acoustics, and bioacoustics. As a result, they were already well-connected and had enough contacts to gather a critical mass for the new community.

2) Although there was no dedicated funding for building the DCASE community, many individual members and institutions had research grants that enabled them to contribute. These contributions included organizing and participating in tasks and workshops, managing the web pages, and writing scientific papers and technical reports.

3) DCASE uses a set of communication tools to efficiently disseminate results and discuss within the community. For example, the DCASE web pages (maintained by Heittola from Tampere University) present the tasks in a clear and coherent fashion, with all editions having a consistent look and level of detail.  The comprehensive presentation of the results allows for system comparison and analysis beyond the mere ranking. Additionally, the web pages include a significant amount of information about datasets and other resources that have been useful in community building. 

4) The community was initially centered around the yearly challenge, which remains an important aspect. The selection of tasks for the challenge over the years has greatly influenced the research directions within the community. The goal in task selection has been to strike a balance between maintaining long-running tasks that attract returning participants and introducing new tasks that open up broader research areas. This outreach to other communities has fostered richer discussions within the community. Task selection and organization have also been crucial for community accessibility. DCASE has ensured that people can participate in the challenge by including entry-level tasks for new participants, including undergraduate students.

5)The DCASE workshop quickly became an established annual event: By 2017, the DCASE Workshop was organized as an independent event, attracting a significant number of participants. The initial workshops in 2016 and 2017 left a positive impression on attendees due to their well-executed organization, including the venue, catering, keynote speakers, and technical program, which helped establish the workshop’s credibility as an independent event.

6) Overall, DCASE has maintained high standards for its tasks and workshops, enhancing its credibility within the scientific community, and attracted participation from neighboring fields. Challenge tasks are required to use open data (as explained in Section II). Participants are required  to submit technical reports with sufficient documentation of their methods; this way, the challenge contributes to openly sharing knowledge within the community and is not just a competition. The challenge promotes originality and open-source submissions by offering awards determined by a jury composed of task organizers.

7) The community strives for diversity and inclusion. Initially, tasks and workshops were organized by representatives from a limited number of institutes. However, in recent years, this has changed significantly. Now, tasks are organized by geographically diverse teams, including representatives from both academia and industry.  
DCASE workshop has also experimented with offering grants to encourage the participation of underrepresented minorities or low carbon footprint transportation to the workshop.

8) The DCASE Steering Group has a central role in the community. One of the group’s responsibilities is to select the challenge tasks and workshop organization teams, and for this, open calls for task and workshop organisation are published to engage the members of the community. 
The Steering Group includes representatives from both academia and industry. It is partially renewed every three years, with attention to maintaining a balance of representation from academia and industry, as well as ensuring gender and geographical equity.

\subsection{Challenges and limitations}

While DCASE community has been developing successfully regarding many aspects, it has its own challenges to overcome.

The challenges rely heavily on datasets. With the choice of openly available datasets, keeping the tasks interesting and evolving requires regularly collecting and curating data. While it can be fairly easy to find new (small) datasets, ensuring that these datasets allow for some continuity with previous tasks and datasets in order to achieve some mid to long-term support is far from obvious. 
Balancing between new and old tasks and opening up to new domains when selecting tasks is influenced by the availability of data. This choice can significantly impact the community’s work, as it involves a trade-off between topics that easily attract participants and those that, while potentially less attractive, could open new avenues in the domain.

The challenge, like any competition, tends to focus attention on the winner of each task. This brings two main drawbacks. First, a significant portion of the work during the challenge or related to the task is aimed at achieving the highest ranking score. This focus can sometimes hinder originality, as innovative solutions may not be as efficient as well-known, finely-tuned approaches. Second, 
to ensure fair comparisons between participants, task organizers must
design some mechanisms to ensure (not always successfully) that everybody follows the rules. These mechanisms can sometimes limit the scientific aspects explored within the tasks.

Organizing an annual workshop is quite demanding, requiring a team of at least five to ten people for each event. One of the community’s challenges is finding new, motivated teams to organize each edition. This is evident in the list of past organizers, which relies on a rather limited pool of individuals who are often already active in the Challenge and the Steering Group.

Communication is a critical aspect in a scientific community. While discussions are very active and fruitful during workshops, they only engage a fraction of the community, as seen when comparing attendance at online and in-person editions. This disparity can be attributed to the various constraints, both time-related and financial, associated with attending in-person workshops. Communication channels were created to overcome this issue, including the DCASE discussion mailing list (since the community’s early stages) and a dedicated Slack channel (since the pandemic). However, these tools are primarily active during the challenge period for announcements by organizers or participant requests. Outside of this period, they remain relatively quiet, falling short of achieving their original goal of fostering ongoing discussions within the community.

\section{Future }

Within ten years, DCASE has established itself as an important research area within the audio signal processing domain. This expansion was greatly helped by the recurring DCASE challenge and the efforts to steer the community towards open science. Yet, this rapid progression relies on a rather limited pool of volunteers, making the community in need of more active participants.

As the community stabilizes after years of growth, one key question is how to keep it lively and attractive? One approach was to open to other communities to foster broader discussions but this must be done carefully to avoid the risk of dispersion in terms of research themes. Managing this remains an open challenge.

Finally, in a world that is rapidly changing, in particular in a context of climate change, but also with the spreading use of machine learning based systems (which are at the core of our research), it becomes essential to question our environmental and ethical impact. DCASE has addressed this recently by including bio-acoustic tasks, monitoring energy footprints, and setting low computational constraints. Yet this is only the beginning of the changes that need to be made to ensure that the work done in our community is beneficial. This is a key aspect we will have to reflect on in the near future.

\bibliographystyle{IEEEtran}
\bibliography{references}

\end{document}